\documentclass[aps,amsmath,prb,twocolumn]{revtex4-1}
\usepackage{bm}
\usepackage{amsfonts}
\usepackage{amssymb}
\usepackage{amsmath}
\usepackage{graphics}
\usepackage{graphicx}
\usepackage[dvipsnames]{xcolor}
\usepackage[colorlinks=true,linkcolor=blue,citecolor=BrickRed]{hyperref}

\def\be{\begin{equation}}       
\def\ee{\end{equation}}
\def\ba{\begin{eqnarray} }
\def\ea{\end{eqnarray} }

\begin{document}

\title{Magnetoenhancement of superconductivity in composite $d$-wave superconductors}

\author{Mauro Schiulaz}
\affiliation{Department of Physics, Boston University, Boston, MA, 02215, USA}
\affiliation{Department of Physics, Yeshiva University, New York City, NY, 10016, USA}
\author{Christopher L. Baldwin}
\author{Chris R. Laumann}
\affiliation{Department of Physics, Boston University, Boston, MA, 02215, USA}
\author{Boris Z. Spivak}
\affiliation{Department of Physics, University of Washington, Seattle, WA 98195, USA}

\begin{abstract}
We study composite $d$-wave superconductors consisting of randomly oriented and randomly distributed superconducting droplets embedded in a matrix. 
In a certain range of parameters the application of a small magnetic field enhances the superconductivity in these materials, while larger fields suppress superconductivity as usual in conventional superconductors.
We investigate the magnetic field dependence of the superfluid density and the critical temperature of such superconductors.
\end{abstract}
\date{\today }

\pacs{}

\maketitle

\section{Introduction}

In general, the superconducting order parameter is a function of two coordinates and two spin indices $\Delta_{\alpha,\beta}({\bf r}, {\bf r}')$. 
Conventional low-$T_c$ superconductors have a singlet order parameter with $s$-wave symmetry which can be described by a complex field $\Delta_{s} ({\bf r}) = \Delta({\bf r}, {\bf r})$. 
Qualitatively, this describes the Bose condensation of Cooper pairs into a zero-orbital-momentum state.
The propagation amplitude of a Cooper pair between two spatial points can be written as a sum of positive partial amplitudes corresponding to different Feynman paths. 
In the presence of a magnetic field, these amplitudes acquire phases and partially cancel one another.  
As a result, $s$-wave superconductivity is suppressed by the magnetic field.
This qualitative picture is consistent with the corresponding solution of the Gor'kov equations \cite{AGD}.

\begin{figure*}
\begin{center}
\includegraphics[width=2\columnwidth]{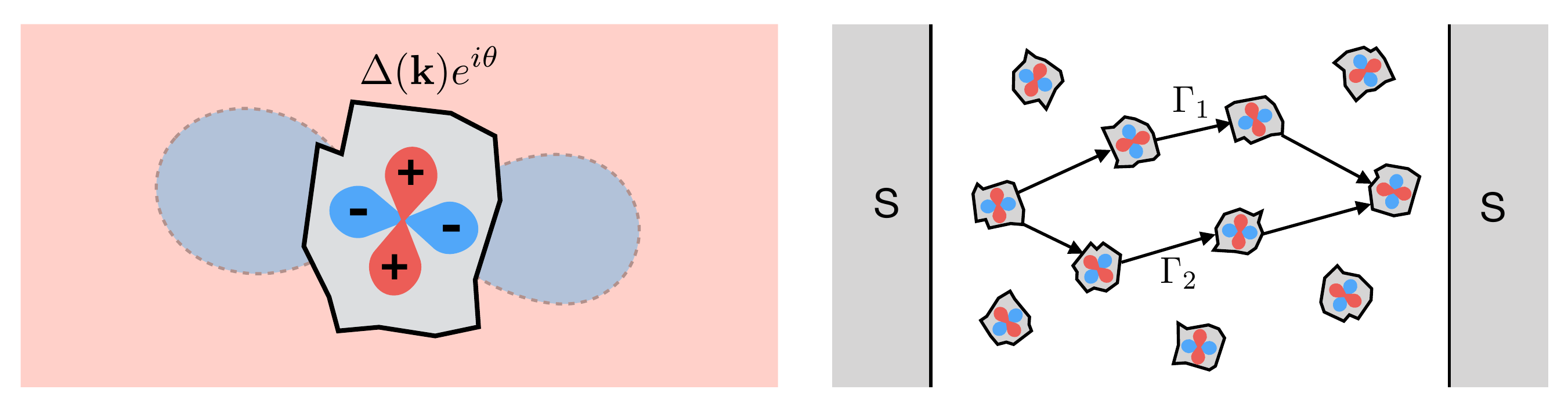}
\caption{Left: Pictorial representation of a $d$-wave superconducting grain and its normal-metallic environment. The grain, colored gray, hosts a nonzero order parameter having the form indicated above the grain. The wave vector dependence of the order parameter is represented by the red and blue rosette on the grain: red corresponds to $\Delta(\textbf{k}) > 0$, and blue corresponds to $\Delta(\textbf{k}) < 0$. The shading outside the grain represents the sign of the anomalous Green's function produced by the proximity effect: red is again positive, and blue is negative.
Right: A granular $d$-wave superconductor sandwiched between two homogenous $s$-wave superconductors, shaded gray. Each individual grain has a randomly oriented order parameter owing to the random orientation of its crystalline axes. $\Gamma_1$ and $\Gamma_2$ indicate two directed paths across the granular system.}
\label{fig:sc_grains}
\end{center}
\end{figure*}

Over the last decades a number of superconductors have been discovered in which the order parameter changes sign under rotation. 
The primary examples are the high-$T_c$ superconductors for which the order parameter has singlet $d$-wave symmetry (see, \emph{e.g.}, \cite{dwave,dwave1}): $\Delta({\bf r}, {\bf r}')$ 
changes sign under rotation by $\pi /2$, and consequently, $\Delta({\bf r},{\bf r})= 0$.  
This means that the Fourier transform $\Delta({\bf k})$ changes sign under a $\pi /2$ rotation as well, as shown schematically in Fig.~\ref{fig:sc_grains}.
Still, the solution of the Gor'kov equation in crystalline materials demonstrates that the application of a magnetic field suppresses superconductivity.

In this paper, we study the magnetic properties of a composite of randomly shaped and randomly oriented $d$-wave superconducting grains embedded in a metallic matrix (see Fig.~\ref{fig:sc_grains}). 
In such systems, the nodes of the order parameter $\Delta({\bf k})$ are locked to the crystalline axes of each grain. 
It is known that the macroscopic properties of such granular materials are distinct from both $s$- and $d$-wave superconductors \cite{oreto,oreto1,KivelsonSpivak}.
Below we show that the application of a magnetic field enhances the superfluid stiffness $N_s$ and the critical temperature $T_{c}$ of such materials in certain parameter regimes.

Granular composites are characterized by the following lengths: 
the typical superconducting grain size $R$, 
the intergrain distance $r_{G}$, 
the elastic electron mean free path in the metal $\ell$, 
the zero-temperature coherence length of the bulk superconductor $\xi_{0}$, 
and the coherence length of the normal metal $L_{T}=\sqrt{D/T}$.  
Here, $D=\ell v_{F}/3$ is the diffusion coefficient, and $v_{F}$ is the Fermi velocity in the metal.   

In the regime where $R, r_G > \xi_0$ and the temperature $T \ll T_c^b$ is smaller than the critical temperature of the bulk superconductor, one can neglect the fluctuations of the modulus of the order parameter and reduce the Hamiltonian to that of a system of Josephson junctions,
\begin{equation}\label{eq:hamiltonian}
H = \frac{\hbar}{2e}\textrm{Re} \sum_{i \neq j} J_{ij} e^{i(\theta_i - \theta_j)}.
\end{equation}
Here, $J_{ij}$ is the Josephson coupling between grains $i$ and $j$, and $\theta_{i}$ is the phase of the order parameter in the $i$th grain.  
Generally, $J_{ij}$ are complex numbers; 
however, in the absence of magnetic field they may be chosen to be real but not necessarily positive. 

Since in random media all spatial symmetries are broken, the anomalous Green's function $F({\bf r}, {\bf r}')$ is an admixture  of $s$, $d$, and higher angular momentum components of the spin-singlet state.  
In the metallic matrix, at distances from the nearest grain greater than $\ell$, only the singlet component survives. 
Thus, in the simplest case where the intergrain distance  $r_G \gg \ell$, the singlet component controls the value of the Josephson couplings $J_{ij}$. 
 
In this diffusive regime and within the mean-field approximation, the $s$ components of the normal $G$ and anomalous $F$ Green's functions satisfy the Usadel equation \cite{Usadel},
\newcommand{\covnabla}{\mathbf{\hat{\nabla}}}
\begin{gather}
\label{usadel}
 \epsilon F_{\epsilon} - \frac{D}{2} \covnabla \left(G_\epsilon \covnabla F_\epsilon - F_\epsilon \mathbf{\nabla} G_\epsilon \right)  = 0,\\
|G_{\epsilon}|^{2}+|F_{\epsilon}|^{2} = 1,\nonumber
\end{gather}
where $\covnabla = \mathbf{\nabla} + 2 e i {\bf A}$ is the covariant derivative, ${\bf A}$ is the vector potential, 
and $F_{\epsilon}( {\bf r})$ and  $G_{\epsilon}( {\bf r})$ are Fourier transforms of the Matsubara Green's functions 
$F({\bf r},{\bf r}, (t-t'))$ and $G({\bf r},{\bf r}, (t-t'))$. 

In the case where $T\ll T_c^b$, the size of the grain is larger than $\xi_{0}$, and the Andreev reflection from its boundary is effective, the boundary conditions for
Eq.~(\ref{usadel}) at the $d$-$n$ boundary were derived in Ref.~\cite{Tanaka}.  
Since the relevant energy for computing the Josephson coupling, $\epsilon \sim  D/r_G^{2}$, is much smaller than the value of the order parameter in the puddles, the boundary condition for $F({\bf r},\epsilon)$ is independent of $\epsilon$ and depends only on the angle between the 
unit vector parallel to the direction of a gap node ${\bf n}_{\Delta}$ and the unit vector ${\bf n(r)}$ normal to the boundary at the point 
 ${\bf r}$ on the surface:
  $F(\epsilon, {\bf r}) = f\{[{\bf n}_{\Delta} \cdot {\bf n(r)}]^2\}$. Here, $f(x)$ is a smooth function, which grows from $f(0) = 0$ to 
 $f(1) \sim 1$.

In the absence of magnetic field ${\bf H}$, a typical spatial distribution of the solution of 
Eq.~(\ref{usadel}) for the anomalous Green's function $F({\bf r},
\epsilon \to 0)$ due to an isolated grain is shown in Fig.~\ref{fig:sc_grains}.  
Red and blue are used to indicate the regions where $F({\bf r},\epsilon \to 0)$ has positive and negative signs, respectively.
The lines where $F=0$ will be of particular interest to us. 

At ${\bf H}=0$,  the phase diagram of the system of $d$-wave droplets embedded in a metal was studied in Refs.~\cite{oreto,oreto1,KivelsonSpivak} .
It has been shown that in the case where the droplets are randomly oriented, the Josephson couplings $J_{ij}$ in Eq.~(\ref{eq:hamiltonian}) are real quantities which can be decomposed as  
\begin{equation}\label{J}
J_{ij} = \eta_{i}\eta_{j} I^{(0)}_{ij} + \eta_{ij} I^{(1)}_{ij}.
\end{equation}
Here, 
\begin{equation}\label{eta}
\eta_{i}=\textrm{sgn }(\int_{s_i} F({\bf r}) d {\bf r})=\pm 1,
\end{equation}
$\eta_{ij}$ are random signs, and the integral in Eq.~\eqref{eta} is taken over the surface $s_i$ of grain $i$.
The positive quantities $I_{ij}^{(0),(1)} $ are randomly distributed on the scales
\begin{equation}
\label{eq:coupling_scale}
I^{(0)}_{ij} \propto \frac{GD}{R^{2}} \frac{R^{d}}{r_{ij}^{d}}\exp(-r_{ij}/L_{T}), \qquad I^{(1)}_{ij} \propto \frac{R^{2}}{r_{ij}^{2}}I^{(0)}_{ij},
\end{equation}
where $G$ is the conductance of a block of the metal of linear size $R$.
Note that the two terms in Eq.~\eqref{J} have different characters.
The first has its sign determined by a product of quantities that depend on the properties of each grain separately, roughly related to the shape of the grains.
Conversely, the sign of the second term is determined by a joint property of the pair of grains $i$ and $j$ (related to the relative orientation of their crystalline axes). 
At large grain concentration where typically $I^{(0)} \ll I^{(1)}$, this problem is a version of the standard model of an XY spin glass~\cite{glass}, while in the opposite limit, the system reduces to the well-known Mattis model~\cite{Mattis}.

In the presence of a magnetic field the Josephson couplings $J_{ij}$ in Eq.~\eqref{eq:hamiltonian} become complex. 
We can generally represent the Josephson coupling at finite $H$ by
\begin{equation} 
\label{eq:josephson_couplings}
J_{ij}({\bf H}) = \pm e^{i \zeta_{ij}} \big| A_{ij} - B_{ij} e^{i \chi_{ij}} \big| I_{ij}.
\end{equation}
Each factor requires some explanation.
The overall scale of the coupling is set by $I_{ij}$ and depends on $r_{ij} / R$, while the sign depends on the specific arrangement of grains $i$ and $j$.
Together, these factors should be thought of as a rewriting of Eq.~\eqref{J}, with $I_{ij}$ being the modulus and the $\pm$ symbol being the sign.
In the limit $r_{ij} \gg R$, $I_{ij}$ maps onto $I_{ij}^{(0)}$, and the $\pm$ sign becomes $\eta_i \eta_j$.
In the limit $r_{ij} \ll R$, $I_{ij}$ maps onto $I_{ij}^{(1)}$, and the $\pm$ sign becomes $\eta_{ij}$.
The remaining factors indicate the effects of a magnetic field.
$\zeta_{ij} = {\bf A}({\bf r}) \cdot {\bf r}_{ij}$, where ${\bf r}_{ij}$ is a vector connecting the centers of grains $i$ and $j$.
The factor $\big| A_{ij} - B_{ij} e^{i \chi_{ij}} \big|$ represents the geometry-dependent proportionality constant from Eq.~\eqref{J}.
$A_{ij}$ captures the positive-weight diffusion paths, and $B_{ij}$ captures the negative-weight paths.
$\chi_{ij} = (H S_{ij}/\Phi_{0})$, where $\Phi_{0}$ is the flux quantum and $S_{ij}$ is the area associated with the diffusion paths, which accounts for the relative phase between positive and negative paths in the field.

We will show that the magnetic field corrections to physical quantities of the system associated with Eq.~\eqref{eq:josephson_couplings} are asymptotically larger than $H^2$ for small ${\bf H}$.
This is the reason why we neglect the quadratic-in-${\bf H}$ suppression of $A_{ij}$ and $B_{ij}$ in Eq.~\eqref{eq:josephson_couplings}.

The value of the area $S_{ij}$ in Eq.~\eqref{eq:josephson_couplings} is also random. 
Its characteristic value  $S$ is not universal. For example, if the diffusion coefficient on the metal in Eq.~\eqref{usadel} does not exhibit spatial fluctuations, $S\sim R^{2}$.

\section{Magnetoenhancement of superconductivity in one dimension}
\label{sec:one_dim}

To illustrate the physical origin of the magnetic field enhancement of superconductivity let us first consider a quasi-one-dimensional case where the droplets are embedded in a metallic wire.  
In the absence of magnetic field the ground state of the system corresponds to $(\theta_{i}-\theta_{j})=0$ if $J_{ij}>0$ and $(\theta_{i}-\theta_{j})=\pi$ if $J_{ij}<0$.
To calculate the macroscopic superfluid stiffness  of the system $\langle N_{S} \rangle$,  we expand Eq.~\eqref{eq:hamiltonian} up to quadratic terms in $\theta_i - \theta_j$ near the ground state.
(We define the superfluid stiffness by the usual equation $\langle {\bf j} \rangle=\langle N_{s} \rangle {\bf \nabla} \theta$, with $\langle {\bf j} \rangle$ being the current density coarse grained on a macroscopic scale.)
As a result, we get the expression
\begin{align}
\label{Ns}
\langle N_{s}(H) \rangle &= \lim_{L \to \infty} \left\langle L\left( \sum \frac{1}{|J_{ij}|}\right)^{-1} \right\rangle \nonumber\\
&=r_{G}\left[\int \frac{p(|J|)}{|J|}d |J|\right]^{-1},
\end{align}
where the sum is taken over neighbor grains, $L$ is the length of the wire, and brackets $\langle\cdot\rangle$ represent averaging over a random distribution of $J_{ij}$ .

At $H=0$, the probability density  $p(|J_{ij}|)$ for the random quantity $|J_{ij}|$ is finite at $|J_{ij}| = 0$.
As a result, the integral in Eq.~\eqref{Ns} diverges  logarithmically,  and the superfluid stiffness is zero. 
Physically, this follows from the presence of arbitrarily weak links in the long wire.

At  $H\neq 0$, the cancellations which produce small $|J_{ij}|$ are less effective because they must cancel in the complex plane. 
The upshot is that $p(|J_{ij}| = 0) = 0$ at finite $H$. 
This cuts off the logarithmic divergence in Eq.~\eqref{Ns}, and we obtain 
\begin{equation}
\label{1dNsEnhancement}
\langle N_{s}(H) \rangle \sim \frac{N_{s}^{(0)}}{|\ln(\phi^{2})|},
\end{equation}
where $N_{s}^{(0)}= \langle |J_{ij}| \rangle$ and $\phi=(H S / \Phi_0)$ is a dimensionless measure of the characteristic flux between grains.
According to Eq.~\eqref{1dNsEnhancement}, the magnetic field enhancement of the superfluid density is nonanalytic, which justifies our neglect of the quadratic-in-$H$ corrections to $J_{ij}$:
physically, the magnetic field suppresses the density of weak links in the long wire.

\section{Magnetoenhancement of superconductivity in $d>1$ dimensions}
\label{sec:dgreater1}

In higher dimensions, the disordered $d$-wave composite superconductor can be frustrated and form a superconducting glass.
This complicates the theoretical analysis.
Below we discuss  several cases where we can nonetheless prove the existence of the magnetoenhancement of superconductivity. 
The suppression of the probability for small couplings $|J_{ij}|$ by a magnetic field is general and independent of dimension, although their effect on the macroscopic superfluid density is dimension dependent.
As we will show, in two and three dimensions the magnetoenhancement is smaller than in one dimension but remains nonanalytic in $H$ (namely, $|H|$). 
We accordingly may neglect all quadratic and higher-order contributions.

\subsection{Magnetoenhancement of superfluid stiffness in the Mattis regime}
\label{sub:mattis}

If the typical intergrain distance is larger than the grain size and the normal metal coherence length, $r_{G}\gg R, L_T$, the second term in Eq.~\eqref{J} can be neglected. 
In the absence of magnetic field, the Hamiltonian~\eqref{eq:hamiltonian} reduces to a Mattis model, for which the random factors $\eta_{i}$ can be gauged out~\cite{oreto,oreto1,KivelsonSpivak}, and accordingly, in Eq.~(\eqref{eq:josephson_couplings}) the $\pm$ sign may be taken to be positive.
In this regime, the phases $\chi_{ij}(H)$ and $\zeta_{ij}(H)$ play different roles. 
We will show that the factors $\chi_{ij}(H)$ inside the modulus in Eq.~\eqref{eq:josephson_couplings} lead to linear-in-$|H|$ enhancement of the superfluid stiffness $N_s(H)$ and critical temperature $T_c(H)$.
On the other hand, the $\zeta_{ij}(H)$ phases produce quadratic-in-$H$ corrections to physical quantities and so we neglect them in the following analysis.
Thus, in this section we take for $J_{ij}(\textbf{H})$ the simpler expression
\begin{equation} 
\label{eq:Mattis_Josephson}
J_{ij}({\bf H}) = \big| A_{ij} - B_{ij} e^{i \chi_{ij}} \big| J_0 e^{-r_{ij} / L_T}.
\end{equation}
We take $A_{ij} + B_{ij} = 1$, with $A_{ij}$ being uniformly distributed in $[0, 1]$ and $\chi_{ij}$ uniformly distributed in $[-\pi |\phi|, \pi |\phi|]$.
Finally, $J_0$ is the characteristic energy scale of the nonexponential front factors in Eq.~\eqref{eq:coupling_scale}. 
Neglecting the variation in $J_0$ is a valid approximation because the disorder in the front factors is subleading compared to that of the exponent.

It is convenient to represent the Josephson couplings in logarithmic variables,
\begin{equation}
J_{ij}=J_{0}\exp(-\xi_{ij}),   
\end{equation}
where 
$\xi_{ij} = \xi^{(0)}_{ij} + \delta \xi_{ij}$,
with
\begin{equation}
	\xi_{ij}^{(0)} = r_{ij}/L_T, \,\,\,\
	\delta \xi_{ij} = - \ln \big| A_{ij}-B_{ij} e^{i \chi_{ij}} \big| .
\end{equation}
This decomposition highlights that the distribution of $\delta \xi_{ij}$ is much narrower than that of $\xi_{ij}^{(0)}$ in the $r_G \gg L_T$ limit.

To calculate the superfluid stiffness of the system $N_s$ at $H=0$, we expand the Mattis Hamiltonian, Eq.~\eqref{eq:hamiltonian}, up to quadratic terms in $\theta_{i}$.
Calculating the superfluid stiffness is then equivalent to calculating the macroscopic conductance of a random resistor network where $\theta_i$ and $|J_{ij}|$ are analogs of the voltage and conductances, respectively.  
In the $r_G \gg L_T$ regime, $|J_{ij}|$ are broadly distributed and we can estimate $N_s$ using percolation theory, as is well known in the context of hopping conductivity~\cite{EfrosShklovskii}.
In this approach, we consider switching on couplings $J_{ij}$ from strongest to weakest, until at a critical value $J_c \equiv J_0 \exp(-\xi_c)$ the network of bonds percolates.
If the couplings are broadly distributed, then the superfluid stiffness $N_s$ is essentially given by $J_c$, analogous to the global conductance of a resistor network being set by the bottleneck with lowest individual conductance.
Reference~\cite{EfrosShklovskii} gives a more detailed discussion.

In the zeroth approximation, where $\delta \xi_{ij} = 0$, we obtain 
\begin{align} 
	\label{eq:unperturbed_superfluid_density}
	\langle N_s^{(0)} \rangle = J_0 r_G^{2-d} \left( \frac{L_T}{r_c^{(0)}} \right) ^{\nu} e^{-\frac{r_c^{(0)}}{L_T}}
\end{align}
where $r_c^{(0)} \equiv L_{T} \xi_{c}^{(0)}$ is the value of $r_{ij}$ at which the network percolates and $\nu$ is the exponent governing the correlation radius of the percolating cluster (e.g., $\nu = 4/3$ in two dimensions, $\nu \approx 0.9$ in three dimensions~\cite{Levinshtein,Sykes}).
See Sec.~5.6 of Ref.~\cite{EfrosShklovskii} for details.

To calculate the magnetic field correction to the superfluid density we use the perturbation theory of percolation theory developed in Ref.~\cite{EfrosShklovskii} (see Sec.~8.3): the first-order correction $\delta \xi_c$ to the percolation threshold $\xi_c$ for typical $\delta \xi_{ij} \ll \xi_c^{(0)}$ 
is given by the average perturbation, $\delta \xi_c = \langle \delta \xi_{ij} \rangle$. Thus,
\begin{align} 
\label{eq:exponent_perturbation}
\delta \xi_c(H) &= - \left\langle \ln{\big| A_{ij} - B_{ij} e^{i \chi_{ij}} \big| } \right\rangle \nonumber\\
&= - \int_{-\pi \phi}^{\pi \phi} \frac{d\chi}{2\pi \phi} \int_0^1 dA \ln{\left|A - (1 - A)e^{i \chi}\right| } \nonumber \\
&\sim 1-\frac{\pi^2}{8} |\phi|.
\end{align}
Thus the superfluid density, which is proportional to $\exp[-\xi_c^{(0)} - \delta \xi_c(H)]$, is enhanced in small magnetic field $\phi \ll 1$:
\begin{align} 
\label{eq:superfluid_density_enhancement}
\frac{\langle \Delta N_s(H)\rangle}{\langle N_s(0) \rangle} \equiv \frac{\langle N_s(H)\rangle - \langle N_s(0)\rangle}{\langle N_s(0)\rangle} \sim \frac{\pi^2}{8}\frac{|H|S}{\Phi_0}.
\end{align}
Note that Eq.~\eqref{eq:superfluid_density_enhancement} does not depend on any details of the percolating cluster such as $\nu$, $\xi_c^{(0)}$, or even dimensionality.
It depends only on having a nonzero probability density for $J_{ij} = 0$, which comes from the fact that the $d$-wave order parameter changes sign as a function of momentum.

The perturbative treatment of the problem which leads to Eq.~\eqref{eq:superfluid_density_enhancement} is valid when the relevant $\delta \xi_{ij}\ll \xi_{c}$. 
On the other hand, as $\phi\to 0$, the main contribution to Eqs.~\eqref{eq:exponent_perturbation} and \eqref{eq:superfluid_density_enhancement} come from intergrain couplings with $|A_{ij}-B_{ij} |\to 0$ for which $\delta \xi_{ij}$ diverges logarithmically. 
The magnetic field suppresses the probability of such events. 
This means that Eqs.~\eqref{eq:exponent_perturbation} and \eqref{eq:superfluid_density_enhancement} are valid if $\phi > \exp(- \xi_{c})$.   
In the opposite limit, at very small magnetic field, the correction to the superfluid stiffness 
$ \langle \Delta N_{s}(H)\rangle / \langle N_{s}(0)\rangle  \sim c \phi^{2}>0$ is  quadratic.  
However, even in this regime, we expect the magnetic field correction to the stiffness is still positive.  
Indeed, at $\phi_c \sim e^{-\xi_{c}}$ the linear and quadratic dependences should match.
This gives us an estimate for the coefficient,
\begin{align}
\label{eq:quadratic_coef}
 	c\sim e^{\xi_{c}}\gg 1.
\end{align}

On the other hand, the conventional negative contributions to the magnetic field dependence of the superfluid density $\langle \Delta N_{s}(H)\rangle/ \langle N_{s}(0)\rangle \sim -a \phi^{2} $ with a coefficient $a$ of order 1.
Therefore, they are dominated by the magnetoenhancement we discuss even for $\phi < \phi_c \sim e^{-\xi_c}$.

\subsection{Numerical simulations of magnetoenhancement in the Mattis regime}
\label{sub:numerics_mattis}

\begin{figure}
\includegraphics[width=0.7\columnwidth]{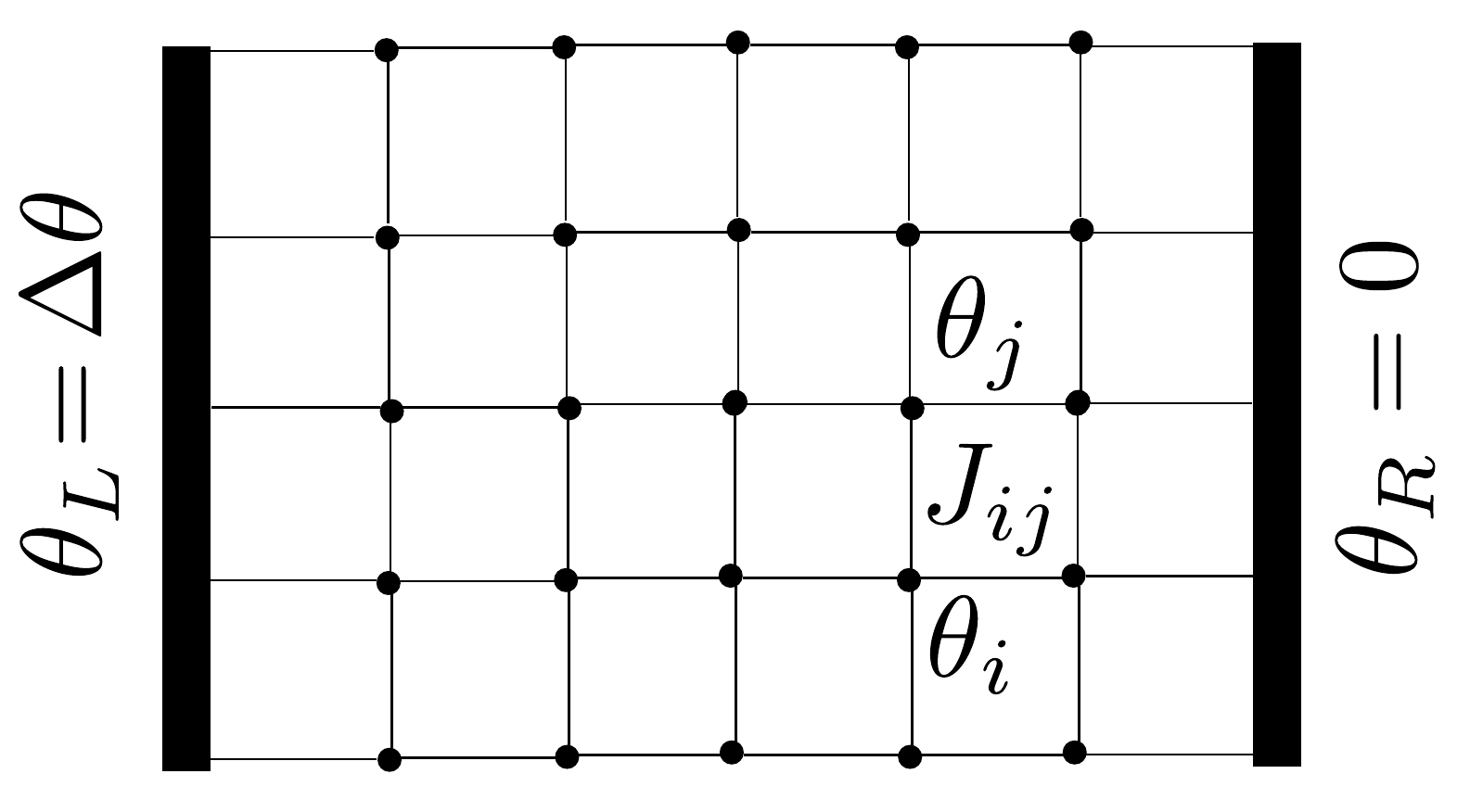}
\caption{The numerical simulations of the Mattis regime are carried out on a square lattice of superconducting grains with random couplings $J_{ij}$. Two large superconducting leads are placed at either end, with $\theta_L = 0, \theta_R = \Delta\theta$, while the system is periodic in the transverse direction.}
\label{fig:network}
\end{figure}

In order to verify the applicability of the perturbative analysis, we simulate the model of Eq.~\eqref{eq:hamiltonian} numerically in the Mattis regime.
We carry out simulations on a regular square lattice of $L \times L$ Josephson coupled grains as in Fig.~\ref{fig:network}. 
At the two boundaries in the $x$ direction, the system is put in contact with a large superconducting reservoir at fixed phase, while the $y$ direction is periodic. 
Since each reservoir is modeled as a single site in contact with all sites on the corresponding boundary, the system has $L^2+2$ sites.

In our simulations, we sample couplings according to the form of Eq.~\eqref{eq:Mattis_Josephson} with $\xi_{ij} \equiv r_{ij}/L_T \in [-W,W]$ uniformly distributed (in units where $J_0 = 1$). 
The parameter $W$ thus represents the typical distance between puddles, in units of $L_T$. 

To compute the enhancement in the superfluid stiffness $N_S$ as a function of the dimensionless magnetic flux $\phi$, we consider the  change in energy due to a small phase difference $\Delta \theta$ (numerically, $\Delta \theta = 1$) applied between the two reservoirs. To leading order, the phases $\theta_i$ at each site $i$ minimize the energy
\begin{equation}
H=\frac{1}{2}\sum_{ij}J_{ij}(\phi) \left(\theta_i-\theta_j\right)^2.
\end{equation}
We find the minimal energy $H^*$ using a quadratic optimization algorithm. The superfluid stiffness is simply given by
\begin{equation}
N_S\propto H^*/\Delta\theta^2.
\end{equation}
We work at disorder strength $0\leq W\leq 8$ and average each measurement of $N_S$ over $N=1000$ random samples.

\begin{figure}
\includegraphics[width=\columnwidth]{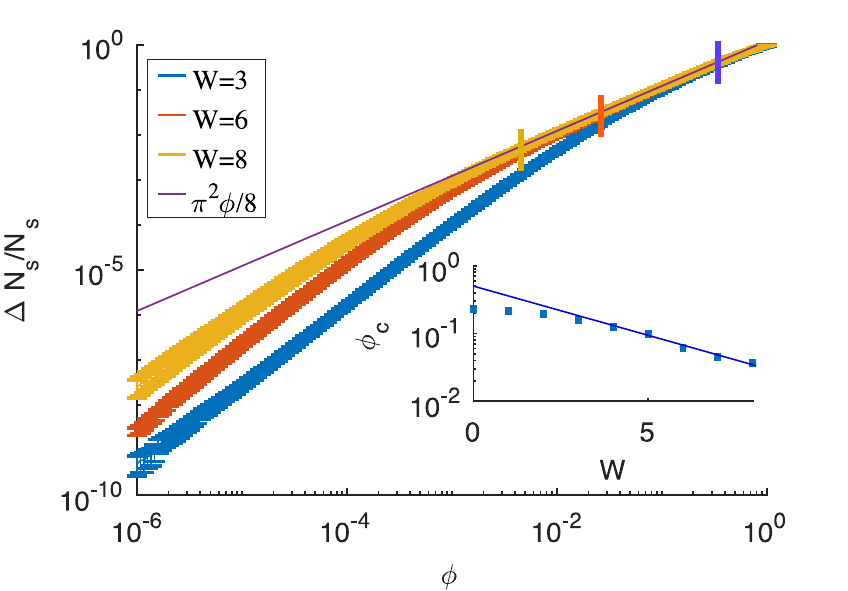}
\caption{
Magnetoenhancement of superfluid stiffness in the Mattis regime of a disordered Josephson network.
Square lattice of linear dimension $L=60$ with $N = 1000$ samples per data point; data for smaller sizes are indistinguishable. The main panel shows the relative enhancement of superconductivity $\Delta N_S/N_S$ as a function of dimensionless flux $\phi$ for several disorder strengths $W$. 
The line $\frac{\pi^2}{8}|\phi|$ is the perturbative prediction of Eq.~\eqref{eq:superfluid_density_enhancement}, which should hold at large $W$ over the range $\phi_c \sim e^{-\xi_c} < \phi < O(1)$. 
At smaller fields $\phi$, the crossover to quadratic behavior is visible.
The crossover point $\phi_c(W)$ is marked by vertical ticks. 
The inset shows the numerically extracted crossover point $\phi_c$ as a function of $W$. The straight-line fit shows the exponential dependence expected at large $W$.}
\label{fig:2d}
\end{figure}

The main panel of Fig.~\ref{fig:2d} shows the dependence of the relative increase of the superfluid stiffness
$\frac{\Delta N_s(H)}{N_s(0)}$
with $\phi$ for several disorder strengths $W$.
Two regimes are clearly visible: for $\phi>\phi_c \sim e^{-\xi_c}$, the behavior is linear and matches the prediction from the perturbative treatment of the percolation theory, $\Delta N_S/N_S=\phi\pi^2/8$.
For $\phi\ll \phi_c$, however, the curves cross over toward quadratic behavior, as expected.

The inset shows the dependence of the crossover point $\phi_c$, at which $\Delta N_S$  becomes linear, on $W$. 
To extract the crossover point numerically, we evaluate the derivative with respect to $\phi$ of the data. Such a derivative grows for small $\phi$ and then saturates to a finite value. 
We estimate $\phi_c$ as the point at which the derivative stops growing. 
The error bars indicate the spacing $\delta\phi$ between two consecutive values of $\phi$.  
The inset compares the numerical data to an exponential fit function $\phi_c\propto\exp(-pW)$, with $p=0.4\pm0.1$.

\subsection{Magnetoenhancement of the critical temperature in the Mattis regime}

One can similarly estimate the change in $T_c$ in a magnetic field.
At the mean-field level, all couplings $J_{ij}$ greater than $T$ are ``rigid,'' so that the phase on the grains connected by such couplings is locked. 
Therefore, the critical temperature may be found by determining when the set of rigid couplings defined by the condition
\begin{equation}
T_{c}(H)<\frac{\hbar}{2e}|J_{ij}|
\end{equation}
percolates.

A similar procedure was applied previously to calculate the critical temperature of disordered ferromagnets~\cite{Korenblit,Kaminski}.
The difference is that in the present case $\xi_c \equiv r_c/L_T$ depends on temperature, so that $T_c$ is determined by the equation
\begin{equation} \label{eq:T_c_unperturbed_equation}
T_c = \frac{\hbar}{2e}J_0 \exp{(-r_c/L_{T_c})}.
\end{equation}
In the absence of $\delta \xi_{ij}$, $r_c = r_c^{(0)}$ is independent of temperature, as previously discussed in Sec.~\ref{sub:mattis}.
Including $\delta \xi_{ij}$ then shifts $r_c / L_T$ according to Eq.~\eqref{eq:exponent_perturbation}; thus, the equation determining $T_c$ is written
\begin{equation} \label{eq:T_c_full_equation}
T_c = \frac{\hbar}{2e} J_0 \exp{(-r_c^{(0)}/L_{T_c} - 1 + \frac{\pi^2}{8} \big| \phi \big| )}.
\end{equation}
Expanding with respect to small $\phi$, we obtain
\begin{equation} \label{eq:T_c_change}
\frac{T_c(H) - T_c(0)}{T_c(0)} \sim \frac{\pi^2}{4} \frac{L_{T_c(0)}}{r_c^{(0)}} \frac{\big| HS \big|}{\Phi_{0}}  .
\end{equation}

The above analysis has a mean-field character in that it neglects fluctuations of the phase between rigid couplings.
However, we expect the conclusions to be correct in the strong-disorder limit ($W \gg L_T$), where all but a vanishing fraction of couplings in the percolating network are much stronger than the putative $T_c$.
Indeed, the authors of~\cite{Kaminski} checked the validity of the percolation theory via Monte Carlo simulations and found that $T_c$ is given by Eq.~\eqref{eq:T_c_unperturbed_equation} up to a factor of order 1.

The magnetoenhancement of $T_c$ behaves analogously to the magnetoenhancement of $N_s$. 
Namely, the linear dependence on $|H|$ applies for fields larger than the previously mentioned exponentially small cutoff $\phi > e^{-\xi_c}$.

\section{Magnetoenhancement in the superconducting glass regime at high temperature.}
\label{sec:glass_regime}

In the superconducting glass regime ($r_G \lesssim R$), the couplings $J_{ij}$ in Eq.~\eqref{eq:hamiltonian} have random signs in the absence of magnetic field.
The frustration this induces at low temperature makes this theoretical problem difficult.
As with spin glasses, most physical properties are out of equilibrium and time dependent.
It is not even clear how to define the superfluid density in general.
Therefore, in this section we restrict ourselves to the case of high temperatures $T \gg (\hbar/2e)|J_{ij}|$, where the system is in the normal state, and show that the superconducting correlation function,
\begin{align}
	\label{eq:corr_function}
A_{kl} = \big< e^{i(\theta_k - \theta_l)} \big> = \textrm{Tr} \left[ e^{i(\theta_k - \theta_l)} \frac{e^{-\beta H}}{Z} \right] ,
\end{align}
is enhanced by a magnetic field.
Here, $H$ is given by Eq.~\eqref{eq:hamiltonian}, $\beta = 1/T$, and $Z$ is the partition function.

This correlation function controls the critical current in a junction composed of two $s$-wave bulk superconductors forming a sandwich around a granular $d$-wave composite (see Fig.~\ref{fig:sc_grains}) in the regime where the temperature is below the critical temperature of the $s$-wave leads.

The sign of the coupling $J_{ij}$ in the glass regime depends on the relative orientation of the order parameter between the two grains. 
We model this dependence by including a factor $\cos{2(\Theta_i - \Theta_j)}$ in Eq.~\eqref{eq:josephson_couplings}, where $\Theta_i$ is the orientation of the positive node of the order parameter on grain $i$.
This factor respects the $d$-wave symmetry of the grains: it retains its sign if either grain rotates by $\pi$ and changes sign if either grain rotates by $\pi/2$.
Furthermore, since the enhancement of the correlation function relies on long-distance universal behavior, as discussed below, we neglect the variation in all other quantities affecting $J_{ij}$ for simplicity. 
This includes the relative phases $\chi_{ij}$ in Eq.~\eqref{eq:josephson_couplings}.
Thus, we model the Josephson couplings as
\begin{equation} \label{eq:nematic_coupling_form}
J_{ij} = J_0 \, e^{i \zeta_{ij}} \cos{2 \big( \Theta_i - \Theta_j \big)},
\end{equation}
where $\zeta_{ij} = {\bf A}({\bf r}) \cdot {\bf r}_{ij}$ and $\Theta_i$ is uniformly distributed in $[-\pi, \pi]$.

The standard high-temperature expansion of Eq.~\eqref{eq:corr_function} gives the correlation function as a sum over paths $\Gamma$ from grain $k$ to grain $l$:
\begin{equation} \label{eq:corr_function_path_sum}
A_{kl} = \sum_{\Gamma} A_{\Gamma} , \quad A_{\Gamma} \equiv \prod_{\langle ij \rangle \in \Gamma} \big( \frac{\pi \hbar \beta}{2e}  J_{ij} \big) .
\end{equation}
The product over $\langle ij \rangle \in \Gamma$ runs over all links along path $\Gamma$. 
Furthermore, since $(\hbar \beta/e) |J_{ij}| \ll 1$, the leading-order terms in the path sum~\eqref{eq:corr_function_path_sum} correspond to \textit{directed} paths.
See Fig.~\ref{fig:sc_grains} for a qualitative example of such directed paths.
In the high-temperature regime, the correlation function decays exponentially at large distance: $\langle \ln{|A_{kl}|} \rangle \sim -r/\Xi(H)$.

\begin{figure}
\begin{center}
\includegraphics[scale=0.6]{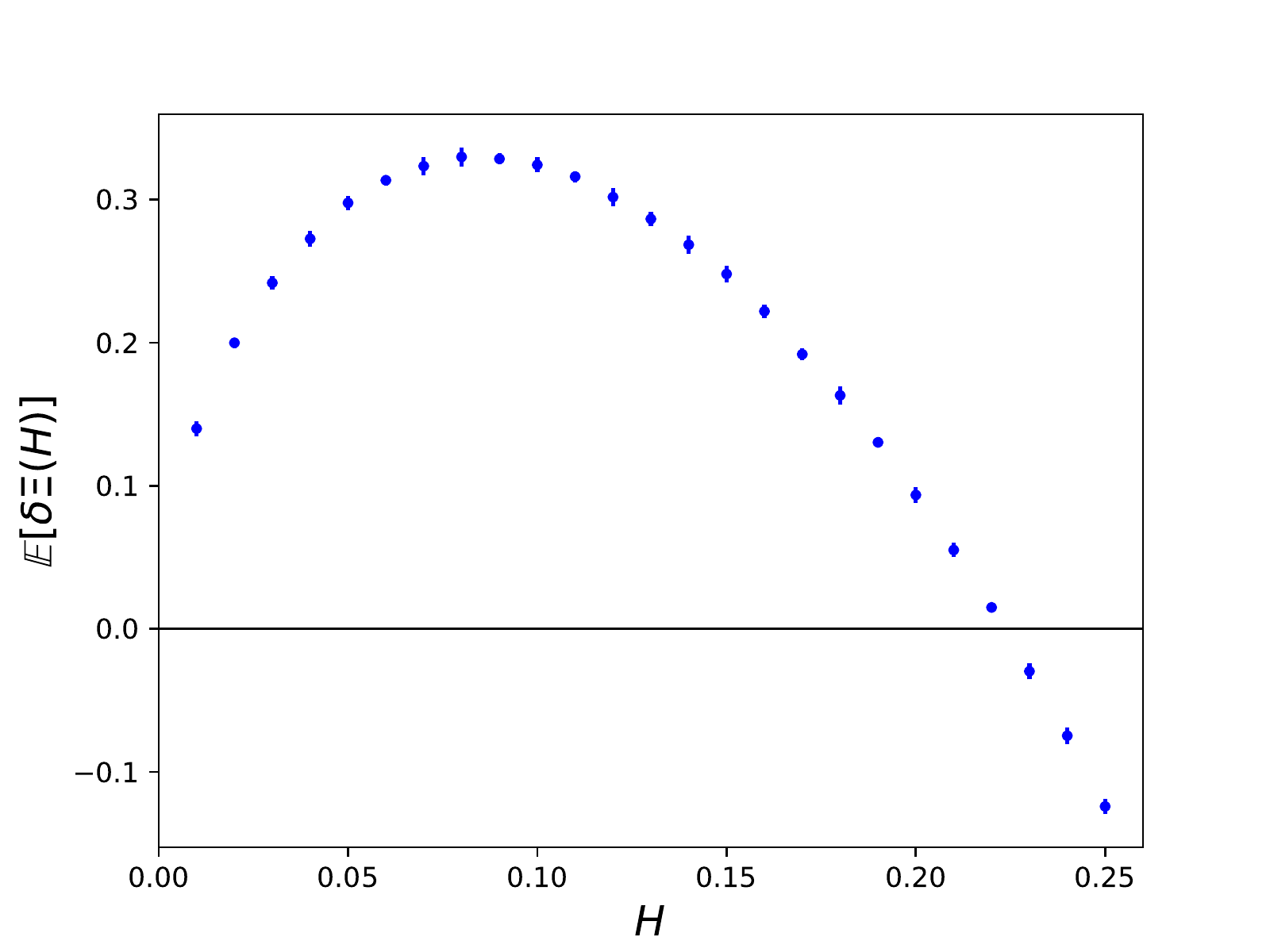}
\caption{The disorder-averaged change in correlation length (in units of lattice spacing) as a function of magnetic field $H$ (in units of flux quantum per lattice plaquette). The sum in Eq.~\eqref{eq:corr_length_expression} is evaluated between opposite corners of a 1000 $\times$ 1000 square lattice. Data for smaller sizes are indistinguishable.}
\label{fig:high_T_magnetoenhancement}
\end{center}
\end{figure}

It follows from Eq.~\eqref{eq:corr_function_path_sum} that
\begin{widetext}
\begin{equation} \label{eq:corr_length_expression}
\frac{1}{\Xi(H)} = \ln{\frac{2e}{\pi \hbar \beta J_0}} - \lim_{r \rightarrow \infty} \frac{1}{r} \ln{\Big| \sum_{\Gamma} \prod_{\langle ij \rangle \in \Gamma} e^{i \zeta_{ij}} \cos{2 \big( \Theta_i - \Theta_j \big)} \Big| }.
\end{equation}
\end{widetext}
We have evaluated Eq.~\eqref{eq:corr_length_expression} using numerical simulations of this model on a 2D square lattice in a uniform perpendicular magnetic field.
The average change in correlation length, $\mathbb{E} [\Xi(H)] - \mathbb{E} [\Xi(0)]$, is plotted as a function of $H$ in Fig.~\ref{fig:high_T_magnetoenhancement}.
At low magnetic field, $\Xi$ increases in a nonanalytic way:
\begin{equation} 
\label{eq:localization_length}
\frac{\Xi(H) - \Xi(0)}{\Xi(0)} \sim \left( \frac{\Xi(0)^2 \, |H|}{\Phi_0} \right) ^{\alpha},
\end{equation}
with $\alpha = 0.59 \pm 0.03$.
This nonanalyticity derives from the statistics of directed paths in disordered media.
Directed path sums have a long history (see Ref.~\cite{HalpinHealy} and references therein), and it is well  known that the governing exponents are universal.
Different microscopic models for the couplings, as long as they include fluctuations, give the same long-distance behavior upon coarse graining.
Thus, we are justified in using the simple Eq.~\eqref{eq:nematic_coupling_form} for $J_{ij}$, and Eq.~\eqref{eq:localization_length} holds.

Indeed, the model defined by Eqs.~(\ref{eq:nematic_coupling_form}) and~(\ref{eq:corr_function_path_sum}) belongs to the same universality class as that used to describe negative magnetoresistance in hopping conductivity~\cite{Shklovskii2,Zhao,Shklovskii3,Ioffe,Baldwin}, so the exponents at small field are the same.
However, at short distances, the model~\eqref{eq:nematic_coupling_form} has much more constructive interference than that in hopping conductivity because the sign of the paths going to grain $i$ are all correlated with the orientation $\theta_i$. 
As a result, at large field where the magnetic length becomes comparable to the ``sign disordering length'' of Eq.~\eqref{eq:corr_function_path_sum} at $H=0$, the magneto-correction to $\Theta$ becomes negative, as observed in Fig.~\ref{fig:high_T_magnetoenhancement}.

\section{Conclusion}
 
We have shown that in certain parametric regimes, the application of a magnetic field leads to nonanalytic enhancement of both superfluid stiffness and the critical temperature in disordered composites of $d$-wave grains embedded in a metallic matrix. 
Heuristically, the magnetoenhancement stems from the suppression of destructive interference between Cooper pairs carrying positive and negative amplitudes in the absence of the field, although the length scale on which this suppression takes place varies between the cases we have considered.

Specifically, we have considered three cases where analytic control is possible.
First, in quasi-one-dimensional wires the macroscopic superfluid stiffness can be inverse logarithmically enhanced from zero by the application of the field. The strength of this effect follows from the suppression of the density of weak nearest-neighbor Josephson couplings by the application of the field.
Second, at $d>1$, where the intergrain distance is much larger than the typical grain size and normal metal coherence length ($r_G \gg R, L_T$) , frustration in the effective system of Josephson couplings is suppressed, and we find that the superfluid stiffness $N_s$ and critical temperature $T_c$ are both enhanced linearly in $|H|$ by mapping onto percolation theory.
Third, in the geometrically frustrated regime ($r_G \sim R$) but at sufficiently high temperature that the Josephson network is disordered, we find that the superconducting correlation length is enhanced with a nontrivial power law $|H|^{\alpha}$, $\alpha < 1$.

We view our results as a proof of principle for magnetoenhancement of superconductivity.
In all of the cases we have presented, our analysis is possible because the system is essentially unfrustrated at $H=0$ and  we can neglect the effects of glassiness and metastability to leading order.
It is an interesting future direction to treat the intermediate, frustrated regimes of this problem directly using more sophisticated numerical techniques.

\begin{acknowledgements}
The authors would like to especially thank S. A. Kivelson for discussions as well as  A. G. Abanov, Y. Cao, B. Gregor, D. Huse, and S. Gopalakrishnan.
C.R.L. acknowledges support from the Sloan Foundation through a Sloan Research Fellowship and the NSF through Grant No. PHY-1656234.
C.L.B. acknowledges the support of the NSF through a Graduate Research Fellowship, Grant No. DGE-1256082.
\end{acknowledgements}

\end{document}